\documentclass[journal]{IEEEtran}
\usepackage[utf8]{inputenc}
\usepackage{cite}
\usepackage{graphicx}
\usepackage{amsmath}
\usepackage{url}
\usepackage{booktabs}
\usepackage{algorithm}
\usepackage[noend]{algpseudocode}
\usepackage{hyperref}

\title{RX-INT\texttt{:} A Kernel Engine for Real-Time Detection and Analysis of In-Memory Threats}

\author{
    \IEEEauthorblockN{Arjun Juneja \\}
    \IEEEauthorblockA{School of Electronics and Computer Science\\
    University of Southampton Undergraduate\\
    aj2g24@soton.ac.uk}
}

\begin{document}
\maketitle

\begin{abstract}
Malware and cheat developers use fileless execution techniques to evade traditional, signature-based security products. These methods include various types of manual mapping, module stomping, and threadless injection which work entirely within the address space of a legitimate process, presenting a challenge for detection due to ambiguity between what is legitimate and what isn't. Existing tools often have weaknesses, such as a dependency on Portable Executable (PE) structures or a vulnerability to time-of-check-to-time-of-use (TOCTOU) race conditions where an adversary cleans up before a periodic scan has the chance to occur. To address this gap, we present RX-INT, a kernel-assisted system featuring an architecture that provides resilience against TOCTOU attacks. RX-INT introduces a detection engine that combines a real-time thread creation monitor with a stateful Virtual Address Descriptor (VAD) scanner alongside various heuristics within. This engine snapshots both private and image-backed memory regions, using real-time memory hashing to detect illicit modifications like module stomping. Critically, we demonstrate a higher detection rate in certain benchmarks of this approach through a direct comparison with PE-sieve, a commonly used and powerful memory forensics tool. In our evaluation, RX-INT successfully detected a manually mapped region that was not identified by PE-sieve. We then conclude that our architecture represents a tangible difference in the detection of fileless threats, with direct applications in the fields of anti-cheat and memory security.
\end{abstract}


\section{Introduction}
\IEEEPARstart{T}{he} increasing sophistication of in-memory code execution techniques presents a problem to modern cybersecurity and anti-cheat methods. Attackers, ranging from state-sponsored actors to cheat developers, have largely migrated from traditional disk-based malware to `fileless' payloads that exist exclusively within the memory of compromised, legitimate processes \cite{crowdstrike2024gtr} \cite{mandiant2024mtrends}. Due to the nature of being `fileless', these threats bypass the primary scanning technique of most antivirus (AV) and security products, allowing them to operate with a high degree of stealth. Manual mapping, a technique where a DLL is loaded without invoking the standard Windows loader \cite{fewer2008reflective}, and module stomping, where the executable code of an already-loaded, trusted DLL is overwritten, are now well documented and commonly used by advanced adversaries and cheat developers.
The challenge in detecting threats of this nature lies in distinguishing legitimate memory operations from malicious ones. Complicated executables, such as games and web browsers, make use of dynamic code generation techniques such as Just-In-Time (JIT) compilation, creating an environment where private, executable memory is not inherently suspicious \cite{chromiumV8Jit}. Existing user-mode memory forensics tools operate with a disadvantage, they must rely on user-mode Windows APIs (such as VirtualQueryEx and ReadProcessMemory) which can be hooked or manipulated by even user-mode rootkits. Their heuristics also often depend on finding structural artifacts like PE headers, which can be deliberately erased by the injector.
Kernel level detectors offer a higher privilege to monitor processes, but they are not without their own weaknesses. A reasonable approach is to periodically scan a process's memory, however, this can create a TOCTOU vulnerability. An adversary can perform a module stomp, execute their payload, and restore the original bytes of the legitimate module in a window of opportunity that is shorter than the polling interval of a periodic scanner. This allows the threat to execute repeatedly while remaining invisible to the detector.
This paper introduces RX-INT, a hybrid, kernel-mode memory forensics system that leverages event-driven triggers to mitigate TOCTOU conditions in in-memory threat detection. It operates from a kernel context and employs a detection engine that combines two distinct methods:
\begin{enumerate}
\item A real-time, event-driven monitor that monitors thread creation to serve as an immediate tripwire for classic injections and, more importantly, as a trigger for the VAD scanner.
\item A stateful VAD scanner that creates a snapshot of a process's memory, including the content hashes of all executable image-backed (MEM\_IMAGE) sections, allowing it to detect modifications. This includes any modifications made within runtime debuggers such as x64dbg.
\end{enumerate}
The synergy between these two components of RX-INT allows it to counter timing-based evasions. A suspicious event from the thread monitor immediately triggers an out-of-band scan from the VAD scanner, mitigating the TOCTOU race condition. While operating from the kernel, RX-INT's primary objective is to counter more advanced evasion techniques initiated from user-mode, establishing a higher barrier of entry for any adversaries than what is required to bypass traditional user-mode scanners, highlighting the techniques to be implemented in existing solutions. RX-INT also introduces a fully in-kernel Import Resolver that programmatically parses the Export Address Tables (EAT) of all modules loaded in the target process. When a suspicious payload is dumped, this resolver scans the raw memory for pointers and automatically generates a report of all resolved API calls, accelerating the reverse engineering process. This can be used alongside a runtime debugger to resolve any imports manually in the case of erased PE headers, or to provide a more complete picture of the threat's behavior.
To validate this approach, we conducted a direct comparison against PE-sieve (v0.4.1), a complex, widely used public memory forensics tool, using a custom-built injection suite alongside some common injectors. RX-INT successfully detected a manually mapped DLL with its PE headers fully erased in memory. Under the same conditions, PE-sieve failed to generate any alerts for either of these techniques.
The contributions via this project are therefore:
\begin{itemize}
    \item A fully in-kernel import resolver that parses Export Address Tables (EATs) to provide automated symbolic analysis of raw memory dumps.
    \item An empirical demonstration that this architecture can detect advanced, evasive in-memory threats that are not detected by other widely-used, powerful tools.
    \item The design and implementation of a hybrid, event-triggered kernel detection architecture that is comparitively resilient to TOCTOU attacks.
\end{itemize}

\section{Background and Threat Model}\label{sec:background}
The efficacy of modern security solutions is increasingly challenged by a class of threats that minimize or entirely eliminate their on-disk footprint. These `fileless' techniques are central to the threat model targeted in this paper, which focuses on an attacker who has achieved code execution on a target system and seeks to inject a payload into a legitimate process to operate stealthily with the goal to modify the legitimate process itself, or to hide malicious code within it. This section details the primary in-memory evasion techniques that RX-INT has been designed to detect, contextualized with adversary behavior.
\subsection{Adversary Goals and Assumptions}
This threat model assumes an adversary in user-mode with at most administrative privileges on a 64-bit Windows system. The adversary's goal is to execute a malicious payload (such as a malicious modification for a game, remote access trojan, or spyware) from within the address space of a trusted process. This is a common form of Masquerading, a sub-technique of Defense Evasion (TA0005) as mentioned by the MITRE ATTACK framework \cite{MITRE_ATTACK_DEFENSE_EVASION}. By operating within a legitimate process's address space, the attacker inherits its trust level and bypasses simple firewalls and monitoring tools. It is assumed the adversary has not yet compromised the kernel (i.e., has not loaded a malicious driver or achieved kernel-level code execution), meaning their actions are initiated from user mode, but they are designed to evade or harm kernel-level detectors.
\subsection{In-Memory Evasion Techniques}
\subsubsection{Manual PE Mapping}
The standard procedure for loading a dynamic-link library (DLL) is the LoadLibrary API \cite{microsoftDLLs}. These functions are heavily instrumented by security products. To bypass this entirely, adversaries can implement their own PE loader. This process, known as manual mapping, involves parsing the PE file format \cite{fewer2008reflective}, allocating a region of virtual memory in a target process with VirtualAllocEx, and manually copying the DLL's sections (.text, .data, etc.) into the allocated block. The injector then performs base relocation and resolves the Import Address Table (IAT) which need to be done manually since it is not reliant on the default Windows API \cite{IredTeamModuleStomping}. A particularly effective variant of this technique involves subsequently erasing the PE headers from the image once it has been copied into memory, which can defeat scanners that rely on finding the IMAGE\_DOS\_SIGNATURE (`MZ') to identify executable modules, part of the reason why PE-Sieve fails. The resulting payload exists as a MEM\_PRIVATE memory region with no clear file backing, making it difficult to attribute. Advanced injectors enhance this by offering options to Clean Data Directories, etc \cite{ghInjector}. This removes all metadata from the in-memory PE image, turning it into a `freeform' blob of code and data that is very difficult to identify with signature-based scans.
\subsubsection{Module Stomping}
Module stomping is a more advanced and stealthy form of process injection. Instead of allocating new private memory, the attacker targets a legitimate, already-loaded DLL within the target process. As detailed by Hammond, this technique involves using VirtualProtectEx \cite{microsoftVirtualProtectEx} to make the legitimate module's executable .text section writable, which is a highly suspicious action~\cite{Orr2019}. The attacker then overwrites a portion of the legitimate code, often the entry point of a known function, with their own malicious shellcode or a trampoline that redirects execution to a different memory region.
This technique is highly evasive for two primary reasons. First, the malicious code is executing from a MEM\_IMAGE memory region, which many security tools inherently trust more than MEM\_PRIVATE memory because it is associated with a legitimate, signed file on disk. Second, extending from the method showcased by F-Secure, a sophisticated attacker can restore the original bytes of the stomped function immediately after their payload executes, defeating periodic memory scanners that check for integrity modifications~\cite{Orr2019}. This creates a critical TOCTOU vulnerability that RX-INT is specifically designed to narrow the window for.
\subsubsection{Code Injection Primitive}
This is the fundamental method of placing and executing code. While classic methods use CreateRemoteThread (internally NtCreateThreadEx), advanced injectors leverage a wide array of alternatives to bypass common API hooks \cite{elastic2023injection}:
\begin{itemize}
\item Thread Hijacking: Instead of creating a new thread (easily detectable), the injector suspends an existing thread in the target, overwrites its instruction pointer (RIP) to point to the malicious code, and then resumes it.
\item QueueUserAPC: A `threadless' injection that queues an Asynchronous Procedure Call (APC) to a legitimate thread. The malicious code is executed when the thread enters an alertable wait state, avoiding the creation of a new thread entirely. Kernel Callbacks \& FakeVEH: Abuse kernel callback functions or set up fake Vectored Exception Handlers (FakeVEH) to hijack the process's control flow in response to system events or deliberately triggered exceptions.
\end{itemize}
\subsubsection{Post-Injection Cloaking Techniques}
After the code is mapped and a thread is executed, injectors can attempt a final layer of techniques to hide the thread itself from analysis tools.
\begin{itemize}
\item Cloak Thread: The thread is created with characteristics that hide it from standard user-mode debuggers.
\item Fake Start Address: The thread's start address in its control structures (TEB/PEB) is pointed to a benign location, while the actual execution begins elsewhere. This is intended to fool scanners that only check the `official' start address.
\item Skip Thread Attach: Manipulate thread flags to prevent standard DLL\_THREAD\_ATTACH notifications from being sent to the process's loaded modules.
\end{itemize}
\subsection{Kernel-Level Threats and Detection Challenges}
While this threat model focuses on user-mode injection, the design of a detector such as this must be informed by the challenges of kernel-level security. The `OnThreadNotify' callback, operating at the kernel level, is designed to catch the creation of threads regardless of user-mode `cloaking' or `fake start address' techniques, providing a reliable, low-level view of process execution that is difficult for a user-mode attacker to subvert. The Windows kernel's internal memory management is organized by a tree of VAD structures, which are opaque to user-mode code \cite{Russinovich2022}, kernel-mode detectors can traverse this tree to build a complete and accurate map of a process's memory layout. However, some research on kernel-level rootkits have shown, even kernel components can be attacked, and relying on a single detection methodology is often insufficient \cite{kruegel2004detecting} \cite{nadim2023kernellevelrootkitdetectionprevention}. Therefore, an ideal detector should employ a multi-layered approach to be effective against an informed adversary \cite{nistDiD}, especially due to the now documented nature of manual-mapping. RX-INT was designed with this principle in mind, combining real-time event monitoring with stateful memory analysis to provide defense. 

\section{System Design and Architecture}\label{sec:design}
\begin{figure}[!t]
\centering
\includegraphics[width=0.9\columnwidth]{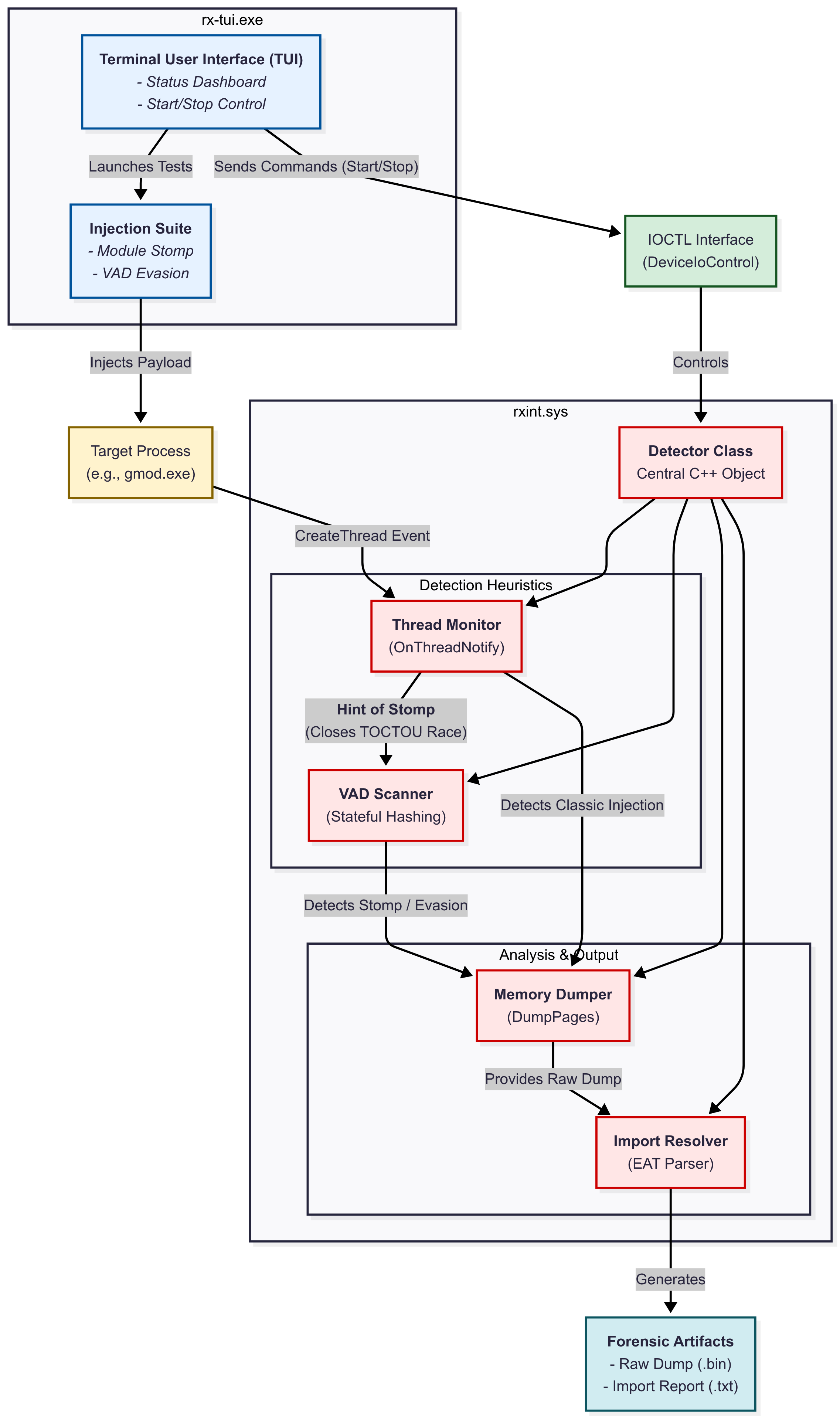}
\caption{Architecture of the RX-INT system, showing the interaction between the user-mode client and the kernel-mode driver components.}
\label{fig_architecture}
\end{figure}
\subsection{User-Mode Client (\texttt{rx-tui.exe})}
The primary interface is the RX-INT TUI. This component is a standalone C++23 executable with zero external dependencies, built on the native Windows Console API for a responsive UI, albeit not being the focus of this project. The TUI's creation had three primary goals:
\begin{itemize}
\item It provides a user with commands to start and stop monitoring on any target process, which is specified by its Process ID (PID). This user-driven control model is crucial for avoiding the false-positives that can arise from attaching to a process during its noisy initialization phase, and also provides more control as a user can attach to and detatch from a process at will.
\item The TUI provides a persistent dashboard that displays the real-time status of the kernel driver (e.g., Idle, Monitoring), the PID of the currently monitored process, and key performance metrics such as the driver's paged and non-paged kernel pool memory footprint. This is how the performance data was collected for the evaluation section.
\item The client contains an `Injection Suite' which serves as a built-in validation and evaluation harness for the kernel driver's detection capabilities. Common injection techniques such as standard manual-mapping were not included, as there are many publicly available tools that can perform these actions \cite{ghInjector}, and the focus of this project is on more advanced techniques that are not commonly tested against. In total, it comes with 5 unique injection methods, each targeting a different aspect of rx-int in order to test for a bypass.
\end{itemize}
\subsection{IOCTL Interface}
Communication between the client and the kernel driver is facilitated through a standard IOCTL model. Upon loading, the driver creates a named device object (\textbackslash\textbackslash Device\textbackslash\textbackslash RxInt) and a corresponding symbolic link (\textbackslash\textbackslash??\textbackslash\textbackslash RxInt) that is visible to user-mode applications, this may act as a detection vector to any adversaries that are trying to hide from RX-INT, however the focus of this project was not targeted towards UM-KM communication. The TUI client uses the CreateFile API to open a handle to this device, and all subsequent commands are sent via DeviceIoControl calls.
This interface is defined by a shared header (ioctl.h) containing a set of custom I/O Control Codes (IOCTLs). The primary IOCTLs allow the client to pass a RXINT\_MONITOR\_INFO structure, containing the target PID and a custom dump path format, to begin a monitoring session, and to send a command to stop the session. Additional IOCTLs are used to query the driver for its current status and memory usage statistics, enabling the TUI's live dashboard.
\subsection{Kernel Driver (\texttt{rxint.sys})}
The core of the system is a Windows (KMDF) kernel-mode driver that performs all detection and analysis. The driver is architected around a central Detector class, which encapsulates all state and logic. To ensure stability and prevent resource leaks in the hostile kernel environment, and due to the ancient nature of the Windows API, the driver makes extensive use of modern C++ RAII (Resource Acquisition Is Initialization) principles. Custom wrapper classes (ProcessReference, SpinLockGuard) provide safe, automatic management of kernel resources such as PEPROCESS object references and KSPIN\_LOCKs. The driver's detection capabilities are divided into two primary subsystems: an event-driven thread monitor and a stateful VAD scanner. Upon a successful detection, the driver is responsible for dumping the suspicious memory region and its automatically generated import analysis report to disk.

\section{Implementation Details}\label{sec:implementation}
To ensure stability and prevent resource leaks, a critical concern in a non-garbage-collected environment like the Windows kernel, the driver makes extensive use of the Resource Acquisition Is Initialization (RAII) paradigm \cite{stroustrup2013cpp}. All critical kernel resources are wrapped in dedicated C++ classes whose destructors guarantee proper cleanup. Key examples include:
\begin{itemize}
\item \texttt{ProcessReference}: This class wraps a PEPROCESS pointer. Its constructor calls PsLookupProcessByProcessId \cite{microsoftPsLookup} to acquire a reference, and its destructor automatically calls ObDereferenceObject \cite{microsoftObDeref}, preventing `zombie process' leaks.
\item \texttt{SpinLockGuard}: Wraps a KSPIN\_LOCK. Its constructor calls KeAcquireSpinLock (saving the old IRQL), and its destructor calls KeReleaseSpinLock. This is used to protect data shared with high-IRQL callbacks, such as OnProcessNotify, and is essential for preventing the IRQL\_NOT\_LESS\_OR\_EQUAL bugcheck.
\item \texttt{ProcessAttacher}: Wraps KeStackAttachProcess \cite{microsoftKeStackAttach} and KeUnstackDetachProcess \cite{microsoftKeUnstackDetach} to safely read memory from a target process's address space.
\end{itemize}
All dynamic memory allocations are routed through a custom tracking system built around a hash table, which allows the user-mode TUI to query the driver's real-time paged and non-paged pool memory footprint via an IOCTL.
\subsection{Stateful VAD Scanner}
The VAD scanner is the primary defense against stealthy modifications to process memory. It operates in a dedicated system worker thread created via PsCreateSystemThread.
\subsubsection{Baseline Creation}
Upon initiation of monitoring for a target process, the worker thread's first task is to establish a comprehensive baseline of the process's virtual memory layout. It traverses the VAD tree by repeatedly calling ZwQueryVirtualMemory, starting from a null base address. It records metadata for every committed memory region into a baseline array stored in paged pool. Crucially, the scanner differentiates between memory types:
\begin{itemize}
\item For MEM\_PRIVATE regions, it records the base address, region size, and memory protection flags. A content hash is explicitly not taken for private memory to keep the baseline's memory footprint minimal, as these regions can be very large and volatile.
\item For MEM\_IMAGE regions with execute permissions (e.g., the .text sections of loaded DLLs), it performs a deeper analysis. The scanner reads the entire section into a temporary non-paged pool buffer using MmCopyVirtualMemory and computes a content hash using the high-speed, non-cryptographic XXH64 algorithm \cite{colletXXHash}.
\end{itemize}
This detailed snapshot, containing both structural information about private memory and integrity information about image memory, serves as the authoritative `ground truth' for the process's legitimate state.
\subsubsection{Change Detection Logic}
After creating the baseline, the thread enters its main detection loop. Periodically (or when triggered by an event), it takes a new snapshot of the process's memory and performs a differential analysis against the baseline. A detection is generated based on the following precise heuristics:
\begin{itemize}
\item  A MEM\_PRIVATE region is found in the new snapshot that has execute permissions and whose address range does not exist in the baseline. This heuristic is effective at detecting classic VirtualAllocEx-based shellcode injections.
\item An executable MEM\_IMAGE region is found whose current XXH64 hash does not match the hash stored in the baseline. This is a high-confidence indicator of a module stomping or inline hooking attack, as the code of a legitimate module should not be modified after it is loaded.
\end{itemize}
\subsection{Event-Driven Trigger and TOCTOU Mitigation}
To mitigate timing-based attacks, the periodic VAD scanner is built with a real-time event monitor. We register a system-wide thread creation callback using PsSetCreateThreadNotifyRoutine \cite{microsoftPsSetCreateThreadNotify}. The callback function, OnThreadNotify, is designed to be a lightweight, non-blocking tripwire.
When a new thread is created in the monitored process, our callback is immediately invoked at APC\_LEVEL. It retrieves the thread's start address via ZwQueryInformationThread and performs a quick analysis of its memory region.
\begin{itemize}
\item If the start address is in MEM\_PRIVATE memory that is not part of a known module (as determined by a PEB walk), it is treated as a direct detection, and the dumping/reporting function is called immediately.
\item If the start address is within a known MEM\_IMAGE region, this is treated as a strong heuristic indicator of a module stomp. Critically, the callback does not perform a full VAD scan itself, as this would be too slow and could lead to deadlocks. Instead, its only action is to immediately call KeSetEvent on a kernel event that the VadScannerThread is waiting on. This awakens the VAD scanner from its sleep state, forcing it to perform an immediate, out-of-band scan.
\end{itemize}
This `hybrid' architecture mitigates the TOCTOU race condition. The detection is no longer limited by a multi-second polling interval but is instead triggered instantly after the CreateRemoteThread call, allowing the VAD scanner to capture the state of the modified memory before an attacker has the chance to restore the original bytes.
\subsection{In-Kernel Automated Import Resolution}
Another contribution of RX-INT is the ability to provide context for dumped payloads, allowing for reconstruction of any of these payload via attaching a debugger to the target or via any `export dumps' RX-INT generates. This is handled by a dedicated ExportResolver class, which builds and maintains a complete snapshot of the target process's export landscape.
The snapshot is built by first enumerating all loaded modules by walking the PEB's InLoadOrderModuleList \cite{chappellPEB}. For each module, the resolver safely copies its PE headers from the target's memory \cite{nebbett2000native}. It then navigates to the IMAGE\_EXPORT\_DIRECTORY and parses the Export Address Table (EAT), the Name Pointer Table (NPT), and the Ordinal Table.
If a function's RVA in the EAT points back into the virtual address range of the export directory itself, the resolver identifies it as a forwarder. It then reads the null-terminated forwarder string (e.g., `ntdll!NtCreateThreadEx') from the target's memory and stores it in the snapshot. The entire snapshot is stored in dynamically allocated paged pool, with a memory-efficient `jagged array' structure to minimize footprint.
When `DumpPages' is called upon any detection, it scans the raw memory dump for any 8-byte value that could be a valid pointer. Each potential pointer is then looked up in the pre-built export snapshot. If the pointer's value matches the absolute address of a known export, the fully resolved symbolic name is written to a companion text report, providing the analyst with immediate, actionable intelligence that would otherwise require a manual debugging session.

\section{Evaluation}
\label{sec:evaluation}
To validate the efficacy, reliability, and performance of RX-INT, we conducted a comprehensive set of experiments. The evaluation was designed to rigorously test the full range of the driver's detection capabilities against a curated suite of advanced, in-memory threats. The primary goals of this evaluation were to: (1) empirically measure the detection coverage for each heuristic against each attack vector, (2) perform a direct, quantitative comparison against a very well-regarded public tool, and (3) quantify the performance and memory overhead of the kernel-mode driver under realistic workloads. These tests were conducted on a Windows 11 24H2 virtual machine with 8GB of RAM and 4 vCPUs, running on a host with an Intel Core i9-13900H CPU and 64GB of DDR5 RAM. The VM was configured to use a dynamically allocated disk image to simulate a realistic environment. 

The primary test target was the 64-bit notepad.exe process to provide a clean, low-noise environment. Additional validation testing was performed on larger processes such as `chrome.exe', `explorer.exe', and `gmod.exe' to assess system stability and basic performance characteristics under realistic conditions. Representative results from these tests are presented in the performance section. An assessment was also conducted during a 10-minute Chrome browsing session, and on a .NET process to determine behavior in a JIT / complex environment. The comparative analysis was performed against PE-sieve (v0.4.1), a widely-used and respected open-source memory forensics tool. For each test, after the injection was performed, a full PE-sieve scan was run on the target process. All PE-sieve scans were run with default parameters (pesieve64.exe /pid <PID>) to ensure a baseline comparison \cite{hasherezade_pesieve}. A detection was considered successful if the tool identified and dumped the injected payload with a non-zero suspicion score. While it can be argued that notepad is not a realistic target for an adversary, it is a clean process that is not known to use any dynamic code generation techniques, and therefore provides a good baseline for testing the detection capabilities of RX-INT. To evaluate false-positives, a separate test was conducted later in the evaluation stage. The tests were run in a controlled environment to ensure repeatability, with each test being executed multiple times to account for any variability in the results. See Table \ref{tab:results}.
\begin{figure}[!t]
\centering
\includegraphics[width=0.9\columnwidth]{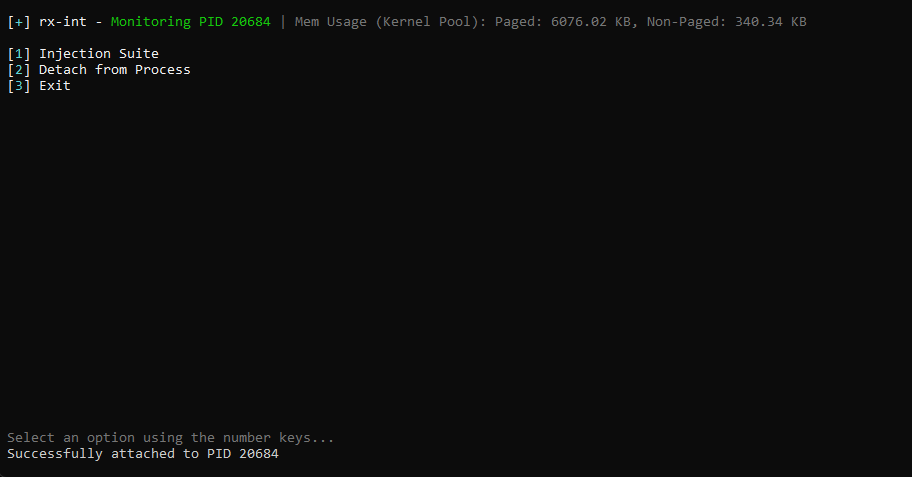}
\caption{The RX-INT TUI client has successfully attached to a target process and is displaying the driver's real-time kernel memory footprint.}
\label{fig_tui_main}
\end{figure}
\subsection{Detection Coverage and Evasion Analysis}
The core of the evaluation was to test the resilience of the hybrid detection model. The results are summarized in Table \ref{tab:results}.

\begin{table*}[ht]
\centering
\caption{Comparative Detection Results between RX-INT and PE-sieve}
\label{tab:results}
\begin{tabular}{@{}llll@{}}
\toprule
\textbf{Attack Scenario} & \textbf{Execution Vector} & \textbf{RX-INT Result} & \textbf{PE-sieve Result} \\ \midrule

Standard Load & \parbox{3.5cm}{NtCreateThreadEx \\ SetWindowsHookEx \\ KernelCallback} & Detected & Detected (Module path visible) \\
\addlinespace
Header Erasure & \parbox{3.5cm}{NtCreateThreadEx \\ SetWindowsHookEx \\ KernelCallback} & Detected & Detected (As detached module) \\
\addlinespace
PEB Unlinking & \parbox{3.5cm}{NtCreateThreadEx \\ SetWindowsHookEx \\ KernelCallback} & Detected & Detected (As detached module) \\
\addlinespace
PEB Unlinking + Header Erasure & \parbox{3.5cm}{NtCreateThreadEx \\ SetWindowsHookEx \\ KernelCallback} & Detected & Error ("Could not read remote PE") \\
\addlinespace
Manual Map + Headers Intact & \parbox{3.5cm}{NtCreateThreadEx \\ SetWindowsHookEx \\ KernelCallback} & Detected & Detected (Dumped as UNMAPPED) \\
\addlinespace
\textbf{Threadless Manual Map + Header Erasure} & \parbox{3.5cm}{QueueUserAPC} & \textbf{Detected} & \textbf{Missed (0 suspicious)} \\
\addlinespace
\textbf{Manual Map + Header Erasure} & \parbox{3.5cm}{NtCreateThreadEx \\ SetWindowsHookEx \\ KernelCallback} & \textbf{Detected} & \textbf{Missed (0 suspicious)} \\
\addlinespace
\textbf{Module Stomping} & \parbox{3.5cm}{NtCreateThreadEx \\ SetWindowsHookEx \\ KernelCallback} & \textbf{Detected} & \textbf{Missed (0 suspicious)} \\ \bottomrule
\end{tabular}
\end{table*}

\subsubsection{Case Study 1, Manual Map with Header Erasure}
\begin{figure}[!t]
\centering
\includegraphics[width=\columnwidth]{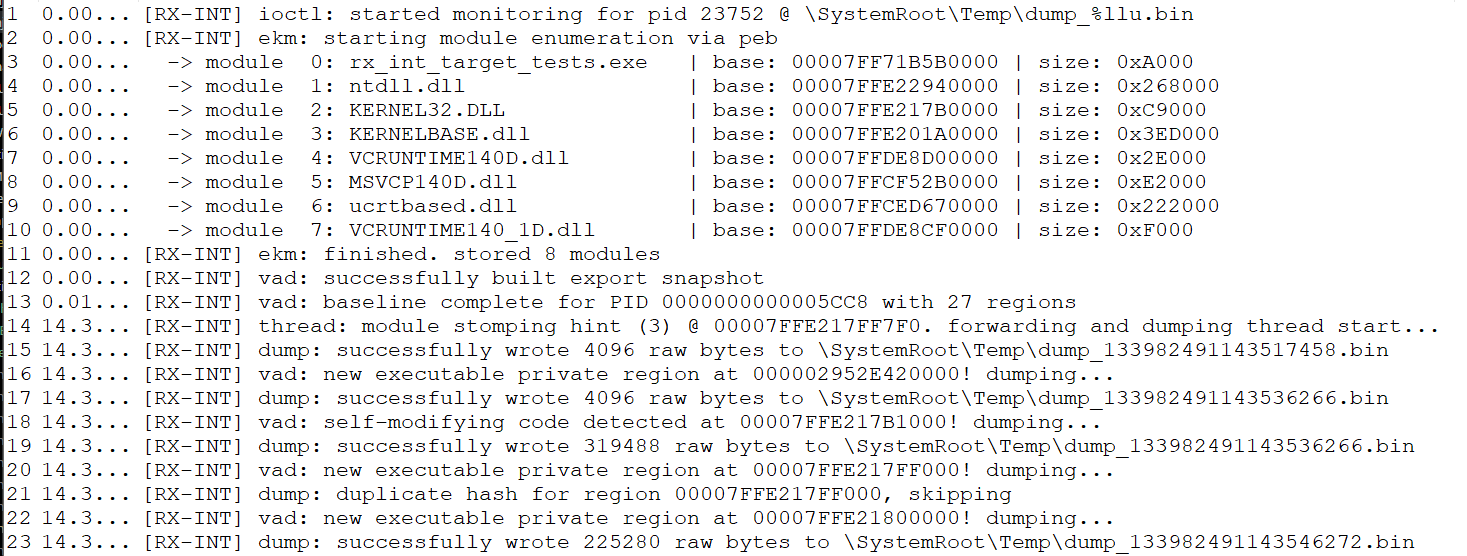}
\caption{Kernel logs from DbgView showing RX-INT's successful detection cascade. The `hint of module stomp' from the thread monitor triggers an immediate VAD scan, which finds and dumps multiple suspicious regions.}
\label{fig_dbgview_modulestomp}
\end{figure}

\begin{figure}[!t]
\centering
\includegraphics[width=\columnwidth]{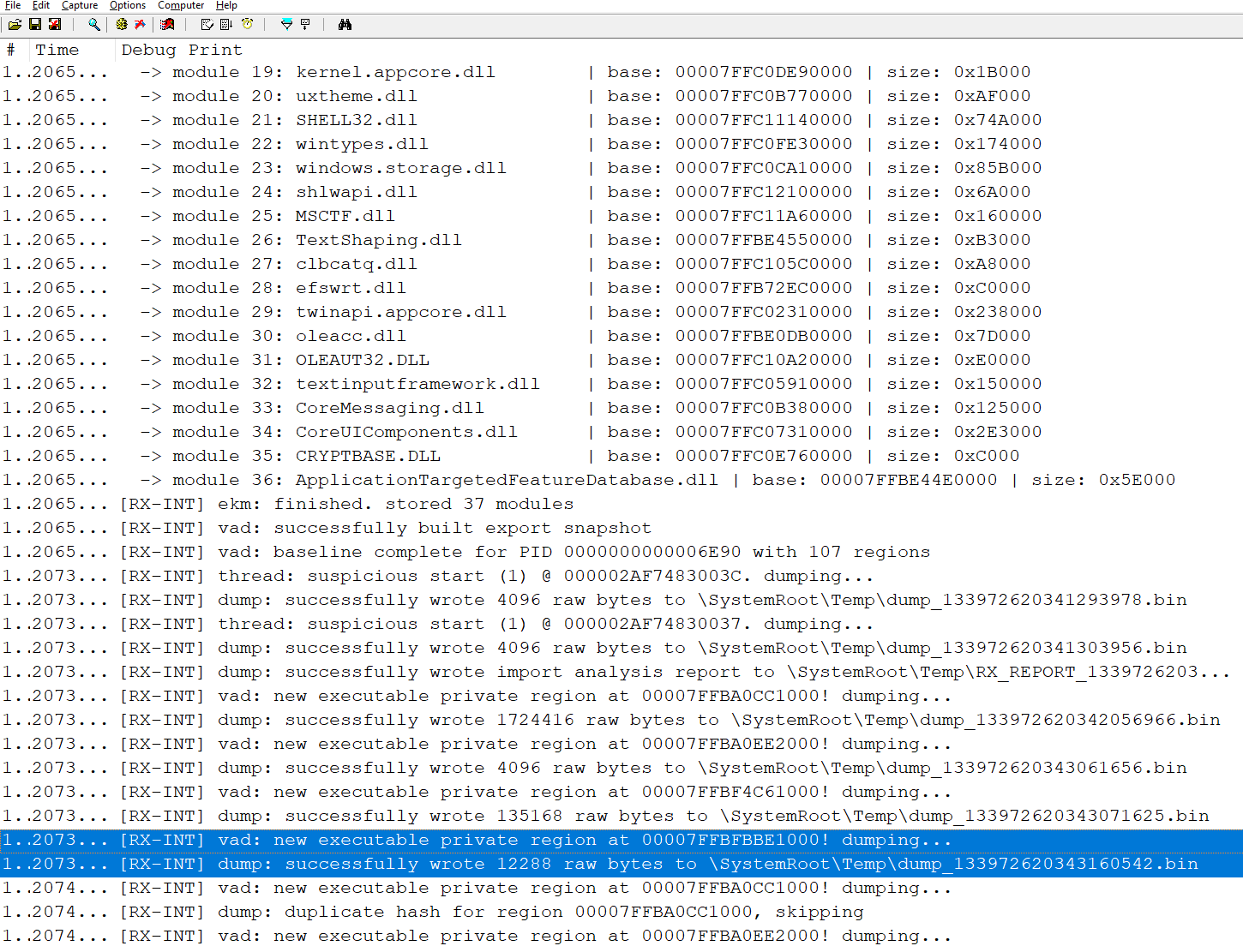}
\caption{Kernel logs from DbgView showing RX-INT's general successful detection.}
\label{fig_dbgview}
\end{figure}

This test simulates an attacker attempting to defeat signature-based scanners. A 64-bit DLL was manually mapped into the target process, and subsequently, the first 0x1000 bytes of the in-memory module, containing the DOS and NT headers, were overwritten with zeroes.
\begin{itemize}
\item \textbf{PE-sieve Result:} A full scan of the process reported no implants. The tool's reliance on finding PE headers to identify modules caused it to miss the payload entirely.
\item \textbf{RX-INT Result:} The stateful VAD scanner detected a new, executable MEM\_PRIVATE memory region that was not present in its baseline. This triggered an immediate alert. The driver successfully dumped the raw memory of the headerless payload and generated an import analysis report. Fig. \ref{fig_ida_rebase} shows the raw dump loaded into IDA Pro, where we have manually rebased the image to its original address (0x7FFBFBBE1000) to begin analysis, confirming the dump's validity. The flow of detection is shown in the kernel logs in Fig.\ref{fig_dbgview_modulestomp} and Fig. \ref{fig_dbgview}, where the VAD scanner detects the new MEM\_PRIVATE region and dumps it immediately after the thread creation event is triggered by the injector.
\end{itemize}

\subsubsection{Case Study 2, Module Stomping}
\begin{figure}[!t]
\centering
\includegraphics[width=0.7\columnwidth]{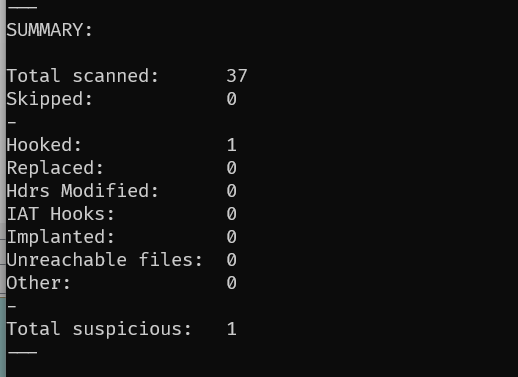}
\caption{The summary output from a PE-sieve scan showing its capability to detect modified modules ("Total suspicious: 1"). However, this detection was not successful against our fast-cleanup module stomping attack.}
\label{fig_pesieve_summary}
\end{figure}

This test simulated an adversary stomping the Beep function in kernel32.dll, executing a payload, and immediately restoring the original bytes.
\begin{itemize}
\item \textbf{PE-sieve Result:} It detected a module (gdi32full.dll) as `suspicious' due to an unrelated false-positive, however was unable to detect any actual module stomping within our target module kernel32.dll. While PE-sieve is capable of detecting modified modules, as shown in the scan summary in Fig. \ref{fig_pesieve_summary}, it was unable to detect the fast cleanup attack. The user-mode tool lost the TOCTOU race.
\item \textbf{RX-INT Result:} The CreateRemoteThread call was immediately caught by the OnThreadNotify callback. This `stomp hint' triggered an instantaneous VAD scan, which detected the content hash mismatch in the legitimate module before the injector could restore the original bytes followed by a successful dump of the modified region.
\end{itemize}

\subsection{False Positive Test on a Complex Process}

To assess RX-INT's resilience to false-positives in high-noise environments, we ran Google Chrome (\texttt{chrome.exe}) under typical usage for a continuous 10-minute session. During this period, the browser was used to open multiple tabs, navigate JavaScript-heavy websites, and load embedded media content. RX-INT reported zero detections throughout the session. 

Given Chrome's extensive use of JIT compilation, dynamic memory allocation, and embedded sandboxed processes, this serves as a realistic baseline for evaluating benign dynamic behavior. While longer runtimes may eventually yield flagged regions due to edge-case heuristics (e.g., highly obfuscated extensions or auto-generated code), no suspicious memory events were observed during the 10-minute monitoring window. 

On the other hand, another test was run on a .NET runtime process with mono, which is known to use dynamic code generation techniques. RX-INT did falsely detect a suspicious MEM\_PRIVATE region, which was confirmed to be benign. 
This false-positive was triggered by the same heuristic that detects `VirtualProtectEx' -based shellcode injections which also leads to a new executable MEM\_PRIVATE memory region being created. While Chrome's V8 JIT engine did not trigger this heuristic in our test window, the .NET CLR JIT did. This result does not invalidate the detection architecture but presents a new research opportunity. Addressing this could have a few solutions, such as analyzing the source of the memory allocation (e.g. whitelisting pages allocated by clrjit.dll) or performing an actual content analysis of the memory region. This highlights the importance of careful tuning and validation in real-world deployment, with RX-INT being a Proof of Concept (PoC) as to what is possible via kernel-level windows internals. RX-INT's heuristics are designed to minimize false-positives while still catching evasive threats.

\subsection{End-to-End Analysis Workflow}
The value of RX-INT is not just in detection, but in its ability to enable a relatively reliable forensic workflow that enables analysis for threats that other tools may not detect. We demonstrate this workflow using the "Manual Map with Header Erasure" test case, which PE-sieve failed to detect.
\subsubsection{Detection and Dumping}
The attack was initiated from the TUI. RX-INT's stateful VAD scanner immediately detected a new, executable MEM\_PRIVATE memory region that was not present in its baseline. The driver automatically dumped two key artifacts: a raw binary dump of the 4096-byte memory region, and an automated import analysis report.
\subsubsection{Initial Analysis in IDA Pro}
\begin{figure}[!t]
\centering
\includegraphics[width=\columnwidth]{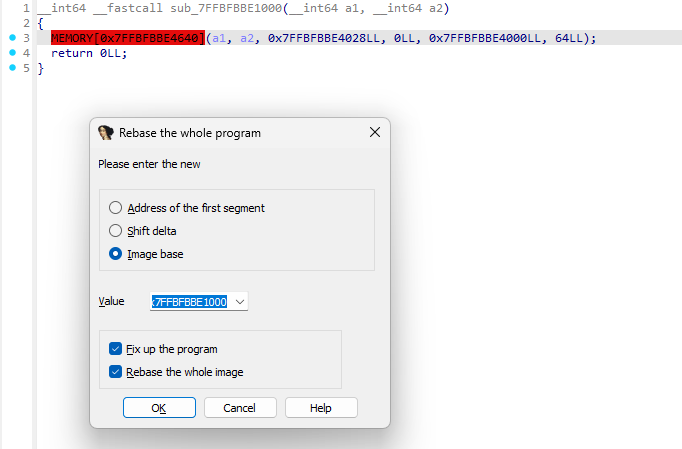}
\caption{The raw memory dump from RX-INT, successfully loaded and rebased in IDA Pro, allowing analysis to begin on the headerless payload.}
\label{fig_ida_rebase}
\end{figure}

\begin{figure}[!t]
\centering
\includegraphics[width=\columnwidth]{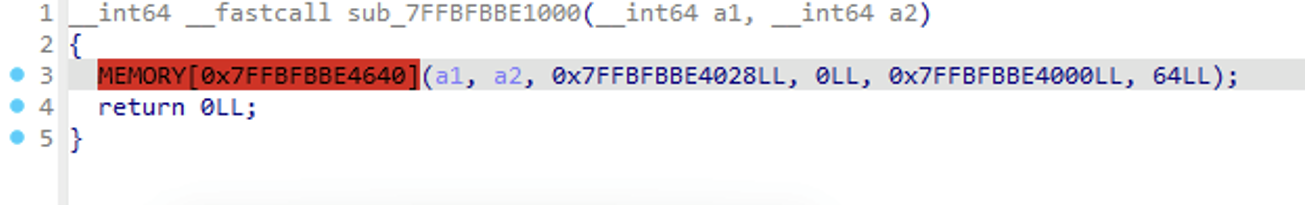}
\caption{IDA Pro's decompilation of the captured payload. The call to an unknown function pointer at MEMORY[0x7FFBFBBE4640] is highlighted.}
\label{fig_ida_pseudo}
\end{figure}

The raw memory dump was loaded directly into IDA Pro. Because the PE headers were erased, IDA initially recognized the file as raw binary data. However, by using the base address logged by RX-INT (0x7FFBFBBE1000) and rebasing the program (Fig. \ref{fig_ida_rebase}), We were able to begin disassembly. IDA's decompilation of a function at the start of the payload, shown in Fig. \ref{fig_ida_pseudo}, revealed a call through a hardcoded function pointer at MEMORY[0x7FFBFBBE4640]. While this confirms executable code, the destination of the call is unknown.
\subsubsection{Symbolic Resolution with the RX-INT Report}
This is where RX-INT's automated analysis can provide a critical advantage. The \_REPORT.txt file generated by the driver can scan the dump and resolve this pointer. In the case that it misses the pointer, the report contains a complete snapshot of the target process's exports, including all loaded modules and their respective Export Address Tables (EATs). If the pointer is not found in the report, the analyst can manually search for it, as mentioned in the next step.
\subsubsection{Live Verification with x64dbg}
\begin{figure}[!t]
\centering
\includegraphics[width=\columnwidth]{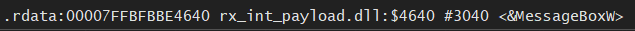}
\caption{Verification in x64dbg. The debugger confirms that the function pointer at the address discovered in IDA and our report (...4640) resolves to user32.dll!MessageBoxW.}
\label{fig_x64dbg}
\end{figure}

To provide the final ground truth, we attached x64dbg to the live, compromised notepad.exe process. We navigated to the address of the function pointer identified by both IDA and our report: 0x7FFBFBBE4640. As shown in Fig. \ref{fig_x64dbg}, x64dbg's symbol engine confirmed the result. It read the pointer at that address and displayed its fully resolved symbolic name: \&MessageBoxW.
This end-to-end workflow, from initial kernel-level detection of a fully obfuscated payload to its symbolic resolution, illustrates the system's ability to provide actionable intelligence on a threat that other tools could not even see, reducing a potentially hours-long reverse engineering task to minutes. This is particularly advantageous in the anti-cheat context, where rapid response is critical to maintaining game integrity and user trust, and detection isnt the only goal, but also providing a way to analyze the threat and understand its behavior.

\subsection{Performance Benchmarks}
To be viable, a security tool must not unnecessarily impact system performance. The RX-INT TUI provides a live view of the driver's kernel memory footprint. As shown in Fig. \ref{fig_tui_main}, the driver maintains a minimal non-paged pool footprint (approx. 340 KB). When monitoring begins, the paged pool usage stabilizes at a mean of 6.07 MB for a complex process like gmod.exe, which is primarily consumed by the one-time creation of the export resolver snapshot. This is a reasonable trade-off for the real-time analysis capabilities it provides.

CPU overhead was measured using the Windows Performance Toolkit. During idle monitoring, rxint.sys consistently consumed less than 0.1\% of total kernel CPU time, having no noticeable impact on the system, as expected. During active detection and polling on `chrome.exe', a brief CPU usage spike corresponding to the VAD scan and memory dump operations was recorded. This spike measured 0.79\% CPU weight. This measurement was conducted on a non-optimized DEBUG build of the driver, which was compiled with optimizations disabled to assist with debugging. The performance of a fully optimized RELEASE build measured a 0.46\% CPU weight, clearly showing that the driver is viable for deployment without adding a significant performance penalty. Considering this is including periodic snapshots of the target process, the performance overhead is negligible and would not be noticeable to the end user.

\section{Limitations and Future Work}
\label{sec:limitations}
While RX-INT demonstrates an advancement in detecting a specific class of in-memory threats, it is not without limitations. This work represents an initial implementation and proof-of-concept for kernel-level fileless malware detection. An obvious limitation remains the evaluation scope: the current testing focuses on demonstrating core functionality against various attack vectors. Comprehensive evaluation against large-scale malware datasets and comparison with multiple tools remains future work. Following this, various offline user-mode tools face similar TOCTOU vulnerabilities as demonstrated with PE-sieve, making direct performance comparison challenging given RX-INT's real-time detection approach.
\subsection{Limitations}
The current in-kernel import resolver parses module Export Address Tables (EAT) directly. While this correctly handles standard and forwarded exports, it does not interpret the modern Windows API Set schema \cite{MSDN_APISets}. The ApiSetMap in the Process Environment Block (PEB) provides a layer of redirection, mapping virtual DLL names (e.g., ext-ms-win-ntuser-message-l1-1-0.dll) to their real host binaries (e.g., user32.dll). As a result, when the resolver encounters a pointer to a redirected API, it correctly identifies the pointer's target within the intermediary module (often ntdll.dll) but does not provide the final, user-friendly symbolic name of the API being called. This represents a gap in the richness of the forensic data provided, which an analyst would currently need to resolve manually with a debugger.
\subsubsection{Vulnerability to Kernel-Mode Threats}
RX-INT is a kernel-mode driver, and as such, it operates at the highest privilege level within the guest operating system (Ring 0). However, it is not immune to attacks from other, malicious kernel-mode components. An advanced rootkit could, in theory, target RX-INT directly. It could hook the PsSetCreateThreadNotifyRoutine or ZwQueryVirtualMemory functions to feed the driver false information, or it could directly tamper with the driver's memory to disable its heuristics or corrupt its baseline data. Defending against such threats would require integrity-checking mechanisms that are outside the current scope of the project. Hooking such functions would also require a bypass of the Windows Driver Signature Enforcement (DSE) and Kernel Patch Protection (KPP) \cite{microsoftDSE}, which is a significant barrier to entry for most adversaries.
\subsection{Future Work: Hypervisor-Based Detection}
The most promising avenue for future work is to transcend the kernel by migrating the entire detection engine to a hypervisor. By leveraging hardware virtualization extensions such as Intel VT-x~\cite{IntelVTX}, a Type-1 hypervisor could be developed to run directly on the hardware, with the entire Windows operating system running as a guest.
This would provide two main capabilities:
\begin{enumerate}
\item The RX-INT detection logic would reside in VMX root mode, making it completely isolated and invisible to all software within the guest, including the guest kernel itself. This would render it immune to kernel-mode rootkits and tampering.
\item \textbf Instead of relying on software-based hashing, we could use Extended Page Tables (EPT) to enforce memory integrity at the hardware level. The hypervisor could mark the physical memory pages corresponding to a module's .text section as read-only in the EPT. An attempt by the guest to write to this memory (as in a module stomp attack) would trigger an immediate, non-maskable EPT\_VIOLATION VM-Exit. This trap would transfer execution to the hypervisor's exit handler, providing a highly reliable reliable, instantaneous, and tamper-resistant detection mechanism with minimal race conditions.
\end{enumerate}
This hypervisor-based approach, often referred to as Virtual Machine Introspection (VMI)~\cite{Garfinkel2003_VMI}, represents a potential evolution of the RX-INT project. This would also allow RX-INT to act as a monitor for the entire system, not just a single process, providing a comprehensive view of all in-memory activity across all processes, and the kernel itself.

\section{Related Work}\label{sec:related_work}
Our work builds upon several foundational concepts while contributing a novel set techniques.
\subsubsection{Offline Forensics}
Frameworks such as Volatility~\cite{Ligh2005_Volatility} are one of the industry standards in memory analysis. They operate on full physical memory dumps of a compromised system and can reconstruct a vast amount of system state, including running processes, loaded modules, and network connections. However, their offline nature means they cannot provide real-time detection or prevention. RX-INT is designed to complement these tools by providing a real-time trigger and memory dump that would serve as the input for a deeper offline analysis with a framework like Volatility.

\subsubsection{UM Scanners}
PE-sieve \cite{hasherezade_pesieve} is an extremely popular tool for the live system scanning of running processes. It uses a set of heuristics to find anomalies, including unbacked executable memory and modified PE headers. As our evaluation demonstrates, however, its reliance on user-mode APIs makes it vulnerable to both TOCTOU race conditions and headerbased obfuscation techniques. RX-INT's primary contribution is overcoming these specific weaknesses by moving the detection logic to the kernel and employing an event-triggered scanning model.

\subsubsection{KM Monitoring}
The use of kernel-mode drivers for security is a common pattern in commercial Endpoint Detection \& Response (EDR) agents. These agents frequently use API hooking (via PsSetCreateThreadNotifyRoutine, etc.) to monitor system events. Our work differs in its focus on the stateful analysis of memory content. While many EDRs focus on behaviors (e.g., a Word document spawning powershell.exe), RX-INT focuses on the state of memory itself, specifically detecting the illicit modification of legitimate code modules, a scope potentially missed by purely behavioral engines. The inclusion of a fully in-kernel EAT-parsing import resolver is a feature typically found only in high-end commercial tools, not in publicly documented research projects.

\section*{Acknowledgment}
We would like to extend gratitude to the open-source reverse engineering community, whose research and tools allowed a project like this to exist. In particular, the extensive work by Aleksandra `hasherezade' Doniec on the PE-sieve memory scanner provided a critical benchmark and motivation for the development of RX-INT's detection heuristics. Additionally, the detailed technical walkthroughs from researchers who publicly document the modern evasion techniques that this work aims to counter served as a foundational resource for developing the project's test harness and threat model. The full project repository can be found at \url{https://github.com/ImArjunJ/rx-int}

\bibliographystyle{IEEEtran}
\bibliography{IEEEabrv,references}

\end{document}